\documentclass{article}

\usepackage{PRIMEarxiv}

\usepackage[utf8]{inputenc} 
\usepackage[T1]{fontenc}    
\usepackage{hyperref}       
\usepackage{url}            
\usepackage{booktabs}       
\usepackage{amsfonts}       
\usepackage{nicefrac}       
\usepackage{microtype}      
\usepackage{lipsum}
\usepackage{fancyhdr}       
\usepackage{graphicx}       
\graphicspath{{images/}}     

\usepackage{listings} 
\usepackage{xcolor}
\usepackage{booktabs} 

\usepackage{hyperref}
\definecolor{lightgray}{gray}{0.9} 
\begin{document}
\title{
	HistoGym: A  Reinforcement Learning Environment for Histopathological Image Analysis
	}
\author{%
	\href{http://zhibo-liu.com}{Zhi-Bo Liu}$^{1}$\thanks{\url{http://zhibo-liu.com}}\quad 
	\textbf{Xiaobo Pang}$^{1}$ \quad 
	\textbf{Jizhao Wang}$^{2}$ \quad
	\textbf{Shuai Liu}$^{1}$\thanks{
		Co-corresponding authors \\
		\quad The code for this project is available at \url{https://github.com/XjtuAI/HistoGym}.
		} \quad 
	\textbf{Chen Li}$^{1}$ \footnotemark[2]\\
$^1$ Xi'an Jiaotong University\\ 
$^2$ First Affiliated Hospital of Xi'an Jiaotong University
}


\maketitle

\begin{abstract}
In pathological research, education, and clinical practice, the decision-making process based on pathological images is critically important. This significance extends to digital pathology image analysis: its adequacy is demonstrated by the extensive information contained within tissue structures, which is essential for accurate cancer classification and grading. Additionally, its necessity is highlighted by the inherent requirement for interpretability in the conclusions generated by algorithms. For humans, determining tumor type and grade typically involves multi-scale analysis, which presents a significant challenge for AI algorithms. Traditional patch-based methods are inadequate for modeling such complex structures, as they fail to capture the intricate, multi-scale information inherent in whole slide images. Consequently, there is a pressing need for advanced AI techniques capable of efficiently and accurately replicating this complex analytical process. To address this issue, we introduce \textit{HistoGym}, an open-source reinforcement learning environment for histopathological image analysis. Following OpenAI Gym APIs, \textit{HistoGym} aims to foster whole slide image diagnosis by mimicking the real-life processes of doctors.  Leveraging the pyramid feature of WSIs and the OpenSlide API, \textit{HistoGym} provides a unified framework for various clinical tasks, including tumor detection and classification. We detail the observation, action, and reward specifications tailored for the histopathological image analysis domain and provide an open-source Python-based interface for both clinicians and researchers. To accommodate different clinical demands, we offer various scenarios for different organs and cancers, including both WSI-based and selected region-based scenarios, showcasing several noteworthy results.
\end{abstract}


\section{Introduction}

Pathology data analysis plays a critical role in the diagnosis, treatment, and prognosis of cancer\cite{Cui2021ArtificialIA}. In clinical practice, pathologists primarily rely on whole slide images (WSIs) for cancer diagnosis\cite{Niazi2019DigitalPA}. However, due to the vast amount of information in WSIs and the limited field of view under a microscope, the process of examining tissue samples reflects the diagnostic approach of pathologists, which is mirrored in teaching practices as well. Pathologists adjust the magnification and navigate through different regions of a slide to make a diagnosis, a process that can be modeled algorithmically by actions such as zooming in, zooming out, and shifting the field of view in digital pathology.


\begin{figure*}[t]
    \centering
    \includegraphics[width=\textwidth]{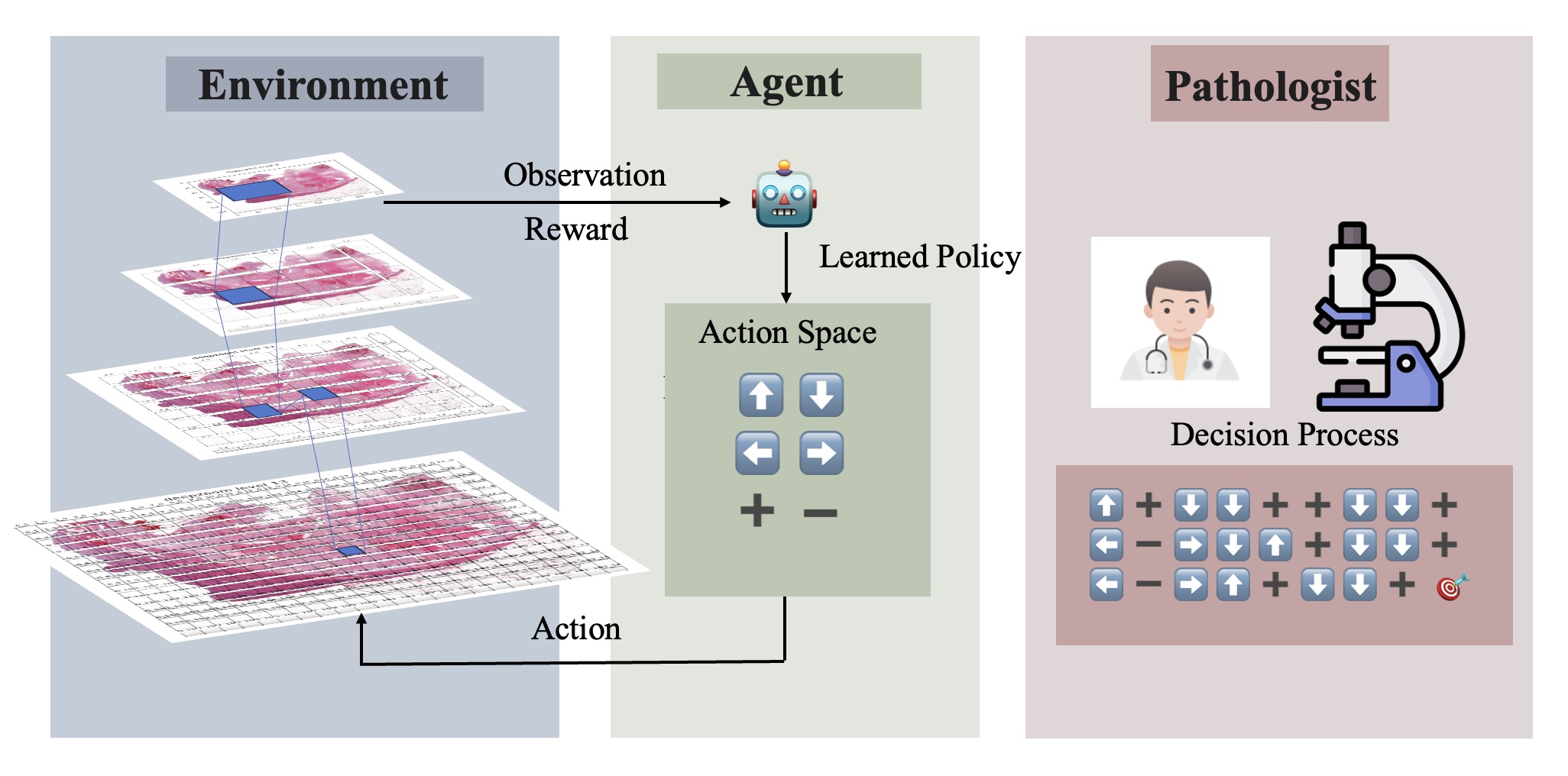}
    \caption{An illustration of the \textit{HistoGym} Environment. This figure illustrates the core components of the HistoGym environment, including the agent’s interaction with the whole slide image (WSI) through a series of actions. The environment leverages tile-based processing and multi-scale navigation, allowing the agent to effectively analyze high-resolution histopathology images.}
    \label{fig:fig1}
\end{figure*}

This paper addresses the question: \textbf{\textit{Is the diagnostic process of doctors better modeled as a decision-making task rather than a classification task?}} We explore whether providing a reinforcement learning (RL) environment to simulate this diagnostic process is a worthwhile endeavor. The motivation for this work is straightforward: to create an RL environment that models the cancer diagnosis process using histopathological data.

Recent advancements in large language models (LLMs), including customized and fine-tuned models, along with the Mixture of Experts model, are increasingly influential in the decision-making process for cancer diagnosis\cite{lu2023visual,huang2023visual}. Multi-modal approaches, incorporating both text and images, are crucial for treatment and prognosis\cite{aflalo2022vl}. Our literature review and studies suggest that exploring the application of RL in this field is both necessary and feasible. Once cancer diagnosis is framed as a decision-making task, explainability, which has long challenged computer vision due to its "black-box" nature, can be addressed more intuitively. By illustrating the diagnostic trajectory, experts can easily interpret the process, despite the time-consuming nature of this approach.

Micro-environmental factors are critical in cancer diagnosis\cite{Binnewies2018UnderstandingTT}. Given the gigapixel scale of WSIs, patch-based methods have been a straightforward solution, with several works introducing hierarchical approaches by combining features from different scales\cite{chikontwe2020multiple,kanavati2020weakly,gao2023semi,shi2024vila}. Reinforcement learning offers a novel way to model these complex micro-environments, capturing information from the organ down to the cellular level without loss\cite{atten-select}.

Although various RL environments exist, few focus on medical image tasks. While some environments, such as those used in gaming\cite{wei2022honor}, robotics\cite{Panerati2021LearningTF}, and autonomous driving\cite{kiran2021deep}, may be too simple or resource-intensive for state-of-the-art algorithms, medical imaging tasks present the opposite challenge—demanding tasks that push the limits of current algorithms.

In this paper, we propose \textit{HistoGym}, a novel open-source RL environment designed to model the diagnostic process in histopathology. Unlike other environments that serve merely as testbeds, \textit{HistoGym} aims to promote the study of treating diagnosis as an RL decision-making problem.

\textbf{Contributions}:
\begin{itemize}
    \item We introduce \textit{HistoGym}, a reinforcement learning environment where agents learn to identify tumor regions at various scales—from organ to tissue to cell level—fully utilizing the pyramid structure of WSIs.
    \item \textit{HistoGym} offers a lossless virtual environment, where agents control their field of view and learn to diagnose slides through actions, mirroring the real-life practices of pathologists. The environment provides configurable feature embeddings and rewards, supporting tasks such as detection and classification.
    \item We evaluate state-of-the-art algorithms across publicly available multi-organ datasets, providing an extensive set of reference results for future research.
\end{itemize}

\section{Motivation and Related Work}

\subsection{Mimicking and Explainability}

Explainability is crucial in medical AI algorithms, impacting research, education, and clinical decision support systems. From an image analysis perspective, two primary levels are considered: patch level and WSI level. At the patch level, techniques such as GradCAM\cite{selvaraju2017grad} and LRP\cite{montavon2017explaining}, as applied to MILPLE\cite{sadafi2023pixel}, are used to highlight key pixels within the patch. At the WSI level, multiple instance learning (MIL) methods are generally categorized into instance-level\cite{chikontwe2020multiple,kanavati2020weakly} and embedding-level\cite{gao2023semi,shi2024vila} approaches, with embedding-level methods being more prevalent due to their ability to handle large data volumes, despite weaker generalization capabilities. In these methods, WSIs are treated as bags, with cropped patches serving as instances. Interpretability is primarily achieved through attention mechanisms that highlight significant patches within the WSI. Some methods combine graph-based approaches with transformers, such as patch-GCN\cite{chen2021whole} and cgc-Net\cite{zhou2019cgc}, to score the entire WSI. However, these methods often fail to explicitly identify salient regions. 

From a visual-textual perspective, recent studies have explored task inference through textual descriptions paired with visual-linguistic similarities, proposing new methods of interpretability\cite{lu2023visual,huang2023visual,aflalo2022vl}. However, these approaches are not well-suited for gigapixel images\cite{li2024llava}. Additionally, in current pathology practices, most paired image-text data is available only at the patch level\cite{huang2023visual,lu2024visual}. This limitation complicates the design of WSI-level interpretable predictive models based solely on patch-level descriptions. Pathology report descriptions are often incomplete, and the image-text data in textbooks is both limited and primarily pedagogical.

\subsection{Computational Efficiency}

Currently, WSI classification is predominantly performed using multi-instance learning (MIL)\cite{song2023artificial}, which necessitates the computation of embeddings for each patch. However, the large number of patches per WSI makes this approach computationally expensive, even when feature extractors are pre-trained on auxiliary tasks\cite{shaban2020context}. Some methods\cite{wulczyn2020deep,wulczyn2021interpretable} attempt to reduce computational burden by randomly sampling patches from a WSI during training; however, this inevitably leads to information loss. Gao et al.\cite{gao2022uncertainty} proposed a prediction rejection strategy based on efficiency uncertainty, which reduces computational burden while efficiently utilizing critical information.

\subsection{Reinforcement Learning Environments in the Medical Domain}

Reinforcement learning algorithms have found success in diverse domains, including board games like Go\cite{silver2016mastering}, video games such as Atari\cite{kaiser2019model} and Honor of Kings\cite{wei2022honor}, as well as in autonomous driving\cite{kiran2021deep}. However, extending their application to the medical domain introduces unique challenges. Historically, RL algorithms have been effective in structured environments, such as Backgammon\cite{tesauro1994td}, Chess\cite{hsu2002behind}, and Go\cite{silver2016mastering}, where the rules are well-defined and easily encoded into algorithms. This specificity may constrain the exploration of RL algorithms in environments that require learning through interaction. The introduction of OpenAI Gym APIs has facilitated the creation of GYM-style environments across various domains, including Gym-preCICE\cite{larsson2020feasibility} for physical AFC applications, Scenario Gym\cite{scott2023scenario} for safe autonomous driving, OpenAI Gym-like\cite{panerati2021learning} environments for robotics applications, and PowerGym\cite{fan2022powergym} for power distribution systems. 

In digital pathology filed, Qaiser et al.\cite{qaiser2019learning} were the first to apply RL to the discrimination of diagnostically relevant regions in WSI, using immunohistochemical (IHC) scores of HER2 as a continuous learning task. Xu et al.\cite{xu2019look} optimized the deep hybrid attention approach with reinforcement learning to achieve high classification accuracy while significantly reducing the number of raw pixels used. Zhao et al.\cite{zhao2023rlogist} subsequently developed RLogist, enabling the agent to explore different resolutions to produce a concise observation path. Building on these advancements, it is crucial to establish a standardized benchmarking environment in pathology. 

\section{Problem Formulation: From Classification to Reinforcement Learning}

In whole slide image (WSI) analysis, traditional machine learning techniques like supervised learning\cite{DLBook}, multi-instance learning (MIL)\cite{Lu2020DataefficientAW, Aguiar2020, Ilse2018AttentionbasedDM, Li2020DualstreamMI}, and reinforcement learning (RL)\cite{sutton-rl-book} offer diverse approaches to address medical imaging challenges. This section outlines the transition from classification-based methods to RL, highlighting the advantages for WSI analysis.

\subsection{Supervised Learning}

Supervised learning is foundational in WSI analysis\cite{litjens2017survey}, where models map input images \( \mathbf{X} \) to labels \( \mathbf{Y} \). Given a dataset \( \mathcal{D} = \{(\mathbf{X}_i, \mathbf{Y}_i)\}_{i=1}^N \), the objective is to minimize the empirical risk:

\begin{equation}
\min_{\theta} \frac{1}{N} \sum_{i=1}^N \mathcal{L}(f_{\theta}(\mathbf{X}_i), \mathbf{Y}_i),
\end{equation}

where \( \mathcal{L} \) is typically binary cross-entropy. This paradigm, though effective, is static and relies heavily on labeled data.

\subsection{Multi-instance Learning}

MIL addresses the challenge of labeling large WSIs  by treating each WSI as a bag \( \mathbf{B}_i = \{\mathbf{x}_{i,j}\}_{j=1}^{M_i} \) of instances (small patches), with the bag labeled positive if any instance is positive:

\begin{equation}
Y_i = \max_{j=1,\dots,M_i} f_{\theta}(\mathbf{x}_{i,j}),
\end{equation}

MIL optimizes bag-level predictions using small patches to generate slide-level representations, eliminating the need for detailed instance labeling and reducing annotation burden\cite{mil97}.

\subsection{Reinforcement Learning for WSI Analysis}

RL introduces a dynamic approach to WSI analysis, framing it as an agent-environment interaction. The WSI is the environment, and the agent's goal is to maximize cumulative rewards through sequential actions. The problem is modeled as a Markov Decision Process (MDP) \( (\mathcal{S}, \mathcal{A}, P, R, \gamma) \):

\begin{itemize}
    \item \( \mathcal{S} \): states, representing the agent's position within the WSI.
    \item \( \mathcal{A} \): actions, such as zooming or panning.
    \item \( P(s_{t+1} \mid s_t, a_t) \): state transitions, deterministic in \textit{HistoGym}.
    \item \( R(s_t, a_t) \): rewards, based on decision accuracy.
    \item \( \gamma \): discount factor.
\end{itemize}

The objective is to learn a policy \( \pi_{\theta}(a_t \mid s_t) \) that maximizes expected cumulative rewards:

\begin{equation}
\max_{\theta} \mathbb{E}_{\pi_{\theta}} \left[ \sum_{t=0}^{T} \gamma^t R(s_t, a_t) \right],
\end{equation}

RL's active learning process offers adaptive strategies for real-time WSI analysis.

\subsection{Transition to RL}

Transitioning from classification to RL represents a shift from passive prediction to active decision-making, where the model interacts dynamically with the WSI environment, optimizing long-term outcomes for improved analysis.

\section{\textit{HistoGym} Environment}

The \textit{HistoGym} environment is a specialized reinforcement learning environment designed for the analysis of whole slide images , providing a comprehensive platform for developing and testing RL algorithms in medical image analysis. This section introduces the core components and functionalities of HistoGym, detailing its integration with the Farama Gymnasium framework, the customization of its RL environment class, and the flexibility it offers in defining various scenarios and observation spaces, thus facilitating advanced experimentation and research in digital pathology.
\subsection{The RL Environment Class}

\textbf{Gym Environment Classes}: The Farama Gymnasium\cite{gymnasium}, originally known as OpenAI's Gym toolkit\cite{gym}, provides a standardized interface for developing and testing RL algorithms in Python. This standardization facilitates the seamless integration of various RL environments, including HistoGym, into established research workflows, enabling consistent and reproducible experimentation.

\subsection{HistoGym Environment Class}

The HistoGym environment is a custom-designed RL environment tailored for the analysis of WSIs. It leverages the OpenSlide\cite{openslide} library for WSI processing and the DeepZoomGenerator for tile generation, allowing the agent to interact with the image data at multiple magnification levels.

Implemented as a subclass of \texttt{gym.Env}, HistoGym adheres to the standard Gym interface, which ensures compatibility with existing RL frameworks such as Stable Baselines3\cite{stable-baselines3} and Tianshou\cite{tianshou}. This design allows the agent to navigate the WSI through a series of discrete actions, including directional movements (up, down, left, right) and zooming in or out.

The \texttt{HistoEnv} class is designed with several key features that are fundamental to its functionality:

\begin{itemize}
    \item State and Observation Space
    
    \item Action Space
    
    \item Reward Mechanism
    
    \item Episode Dynamics
\end{itemize}

These components are discussed in detail in the following subsections, where we explore how each element contributes to the overall effectiveness of the \texttt{HistoEnv} framework.

\subsection{RL Framework}

To effectively implement and experiment with reinforcement learning algorithms in the HistoGym environment, we employ several state-of-the-art RL frameworks, including \texttt{stable\_baselines3} \cite{stable-baselines3}, \texttt{Tianshou} \cite{tianshou}. These frameworks offer a range of pre-built algorithms and tools for training and evaluating RL agents, facilitating seamless experimentation and benchmarking across different RL strategies within \textit{HistoGym}.

\subsection{Environment Complexity \& Scenario of Different Levels}

\begin{figure}[t]
    \centering
    \includegraphics[width=0.75\textwidth]{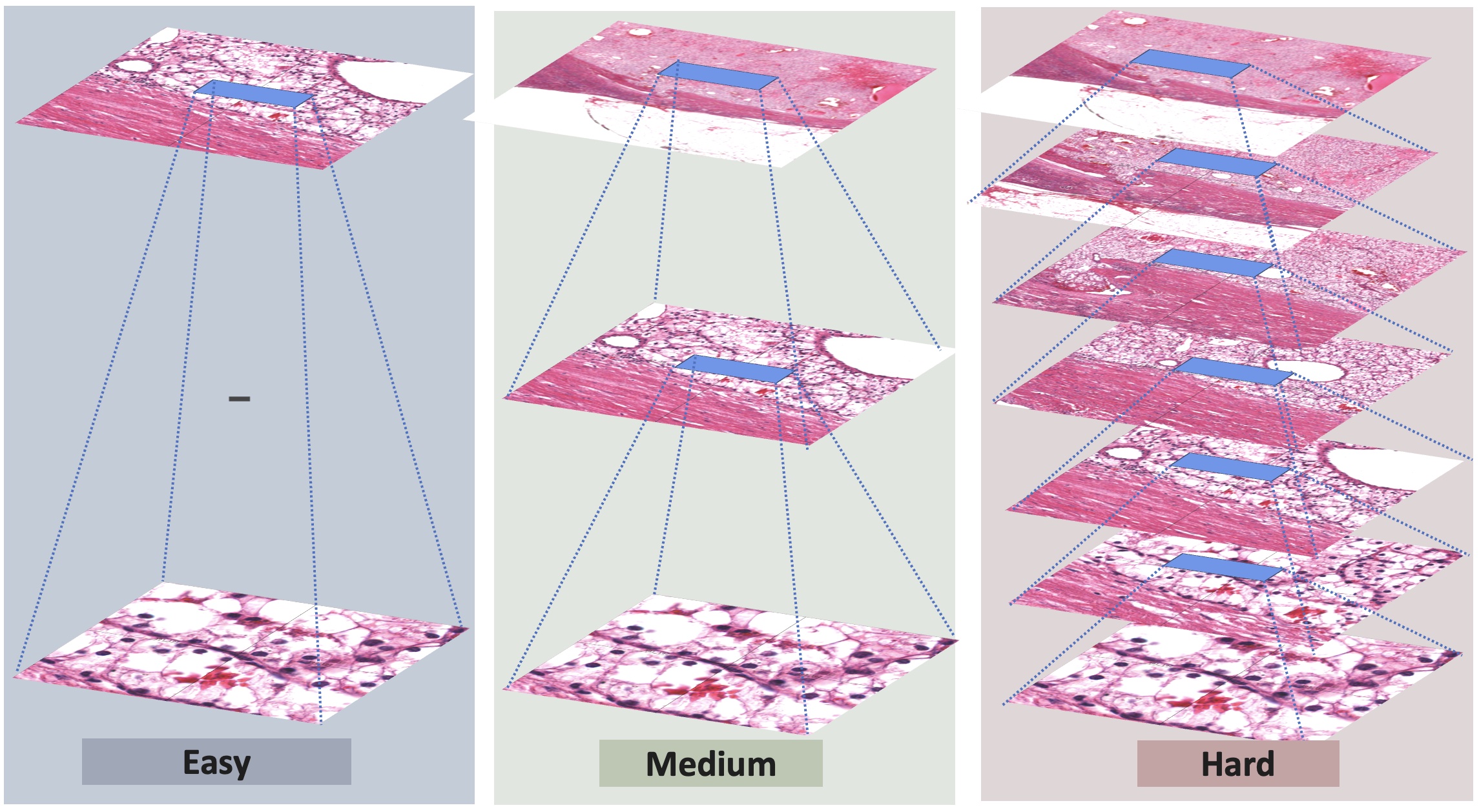}
    \caption{\textit{HistoGym} Environment Complexity}
    \label{fig:fig2}
\end{figure}

HistoGym supports multiple scenarios with varying levels of detail, enabling researchers to tailor experiments to specific research questions or application needs. These scenarios range from basic setups focused on simple navigation and scanning tasks to more complex configurations that require sophisticated decision-making processes as shown in Figure \ref{fig:fig2}. Designed to facilitate academic research, these scenarios provide a controlled environment where hypotheses about RL strategies in WSI analysis can be rigorously tested and validated.

For example, one scenario might involve the agent learning to prioritize regions of the WSI that are more likely to contain pathological features, while another might require the agent to optimize the scanning process to minimize the time required for diagnosis. The flexibility in scenario design allows researchers to explore a wide range of challenges and solutions within the RL framework.

\subsection{State \& Observations}

In the HistoGym environment, the \textit{state} represents the complete set of information returned by the environment after an action is taken. This includes data such as the agent's current \(x, y\) position on the slide, the zoom level \(z\), the type of tissue being analyzed, and the current pixel data from the WSI.

An \textit{observation}, on the other hand, refers to any transformation of the state that serves as input to the control algorithms. Observations are critical for guiding the agent's decision-making process, and different representations can be employed depending on the research focus. We propose three distinct observation representations:

\begin{itemize}
    \item \textbf{Pixel Space}: This representation utilizes raw pixel data normalized from \([0, 255]\) to \(R \in [0, 1]\), making it the most intuitive but computationally intensive method. Pixel space observations are valuable for research that involves fine-grained analysis, as they provide a direct view of the WSI.
  
    \item \textbf{Feature Space}: In addition to high-dimensional RGB pixels, low-dimensional embedding features are also supported. We leverage pretrained feature extractors such as ResNet \cite{resnet} and CLAM \cite{clam} to generate compact feature representations:
        \begin{itemize}
            \item \textbf{ResNet Embedding}: Feature maps generated by ResNet\cite{resnet} encode structural and textural information about the tissue, capturing essential characteristics necessary for accurate analysis.
            \item \textbf{CLAM Embedding}: The CLAM\cite{clam} embedding provides a compact vector summarizing various aspects of the WSI, including tissue type, cell density, and specific biomarkers, making it suitable for scenarios requiring more abstract representations.
        \end{itemize}
\end{itemize}

\subsection{Action Space}

The HistoGym environment defines a versatile action space, categorized into two primary types: discrete and continuous actions.

\begin{itemize}
    \item \textbf{Discrete Actions}: These include standard movement actions (e.g., up, down, left, right) within the same zoom level, as well as actions to zoom in or out. In the initial version of the environment, actions were strictly constrained, requiring the agent to follow discrete steps. For example, after moving in a direction, the agent scans the adjacent region, and after zooming in, it focuses on the next level of detail within the WSI.
    
    \item \textbf{Continuous Actions}: To overcome the limitations of discrete actions and better explore the expansive state space, we extended the action space to include continuous actions. This allows the agent to perform more granular adjustments, such as smoothly zooming in or out and making fine-grained movements across the slide. Continuous actions are particularly useful in optimizing the scanning process and improving the agent's overall performance \cite{mnih2016asynchronous}.
\end{itemize}

\subsection{Episode Dynamics}

An episode in the WSI analysis task is defined by the sequence of actions taken by the agent from the start of the slide analysis to the completion of the task. The episode concludes when the agent successfully identifies the target region (e.g., a cancerous area) or when a predefined scanning limit (e.g., maximum number of steps) is reached. The agent's performance across different episodes can be analyzed to assess its learning progress and the effectiveness of the RL strategies employed \cite{silver2016mastering}.

\subsection{Rewards}

\textit{HistoGym} supports both sparse and dense reward configurations, each tailored to different analysis tasks and tissue types. The reward function is managed by the \texttt{Coor.check\_overlap} module, which calculates the reward score based on the overlap ratio between the agent’s bounding box and the target region within the WSI. Sparse rewards are given for achieving significant milestones, such as identifying the target region, while dense rewards provide continuous feedback based on the agent’s proximity to the target. These configurations allow researchers to explore different reward structures and their impact on the agent's learning process.


\section{Example \& Experiments}

\subsection{Python Example}
The HistoGym environment follows the widely used Farama Gymnasium API \cite{gymnasium}. Below we show an example code that runs a random agent on our environment. We provide an example for using our environment in Listing 1.For more detail, please refer to Appendix and the github documentation page.

\subsection{Experimental Setup}
In this section, we detail the experimental validation of state-of-the-art deep reinforcement learning algorithms within the \textit{HistoGym} environment. We specifically focus on two widely recognized algorithms: Proximal Policy Optimization (PPO)\cite{ppo} , a policy gradient method known for its robustness, and Ape-X DQN\cite{Horgan2018DistributedPE}, a distributed version of the Deep Q-Network(DQN) \cite{dqn} that leverages modern advances in deep learning for enhanced performance.

All experiments utilized ResNet-based feature representations as inputs for observation and discrete actions.  The experiments were conducted using the \texttt{stable-baselines3} \cite{stable-baselines3} library, a widely used framework that ensures consistency and reproducibility in reinforcement learning research. For full details on the training setup, including architecture and hyperparameters, we refer the reader to our GitHub repository .

\begin{lstlisting}[caption={Example Code for Running a Random Agent in \textit{HistoGym}}]
import numpy as np
import numpy as np
from gym_histo import HistoEnv

# Initialize Arguments
img_path ='/path/to/wsi.tif'
xml_path = '/path/to/annotaion.xml'
tile_size = 128 
result_path = './results'

env = HistoEnv(img_path, xml_path, tile_size, result_path)
obs = env.reset()

done = False
while not done:
    action = env.action_space.sample()
    obs, reward, done, info = env.step(action)
    print('action=', action, 'info=', info, 'reward=', reward, 'done=', done)
    env.render(mode="human")
    if done:
        print("Episode Terminated", "reward=", reward)
        break
\end{lstlisting}

\subsection{Impact of Environmental Complexity}

The experimental results, as detailed in Table \ref{tab:tab1}, underscore the profound influence of environmental complexity on both training dynamics and the performance of agents in detecting cancerous regions within WSIs. We quantify environmental complexity by the number of levels in the WSI, utilizing \texttt{openslide} to define the easy, medium, and hard settings, corresponding to 3, 5, and 7 levels, respectively. As shown in Figure \ref{fig:fig2}, increasing the number of levels reduces the magnification between layers, thereby expanding the search space and increasing the difficulty of the task.

In our study, we also investigated the performance of agents under discrete and continuous action spaces. Notably, our findings reveal that as environmental complexity surpasses the simple level, vanilla PPO and DQN models struggle to converge when operating with continuous actions. This divergence highlights a critical area for further research, as it suggests that these models may be ill-suited for handling complex, continuous decision spaces in WSI analysis.

In scenarios characterized by simpler environments and discrete action spaces, both PPO and Ape-X DQN were able to successfully identify cancerous regions within a reasonable number of timesteps. However, as the complexity increased, neither DQN nor PPO achieved convergence. This performance degradation likely stems from the expanded search space and the challenges associated with sparse reward signals in more intricate environments. The inability of these models to converge under high complexity conditions indicates that current approaches may require substantial modifications or novel strategies to effectively navigate and analyze complex WSIs."

\begin{table}[h]
\centering
\caption{Performance under different environmental complexities.}
\begin{tabular}{lcccc}
\toprule
\textbf{Env. Complexity} & \multicolumn{2}{c}{\textbf{Reward(PPO)}} & \multicolumn{2}{c}{\textbf{Reward(DQN)}} \\
                & \textbf{Cont.} & \textbf{Discrete} & \textbf{Cont.} & \textbf{Discrete} \\
\midrule
Easy & 0 & 18.4 & 0 & 8.7\\ 
Medium  & 0 & 11.3 & \textless{}0 & \textless{}0 \\ 
Hard & \textless{}0 & \textless{}0 &\textless{}0 & \textless{}0 \\

\bottomrule
\end{tabular}
\label{tab:tab1}
\end{table}

\subsection{Representation Learning from Raw Observations}
An intriguing research direction involves training reinforcement learning agents directly from raw pixel data, a method that has demonstrated success in simpler environments like Atari \cite{dqn}, but remains particularly challenging in more complex domains such as histopathological image analysis. Within the \textit{HistoGym} environment, we compared the effectiveness of using raw pixel data against more abstract feature representations, such as those extracted by ResNet and CLAM, as detailed in Table \ref{tab:tab2}. Our findings suggest that simultaneously learning the policy alongside the feature extractor is demanding and introduces instability into the model, which warrants further investigation.
\begin{table}[h]
\centering
\caption{Comparison of representation learning approaches in \textit{HistoGym} using  CAMELYON16}
\begin{tabular}{lcc}
\toprule
\textbf{Representation} & \textbf{Reward(PPO)} & \textbf{Reward(DQN)}\\  
\midrule
Pixels   & \textless{}0 & \textless{}0 \\ 
ResNet Features & 13.2 & 8.3 \\ 
CLAM Features  & 16.2 & 8.5 \\ 
\bottomrule
\end{tabular}
\label{tab:tab2}
\end{table}

\section{Conclusion}

In this paper, we introduced \textit{HistoGym}, an open-source reinforcement learning environment tailored for histopathological image analysis. By reframing the diagnostic process as a decision-making task, \textit{HistoGym} provides a novel platform for exploring the application of RL in medical imaging, effectively mirroring the real-world practices of pathologists. This environment not only addresses key challenges in WSI analysis, such as handling high-dimensional data and integrating multi-scale information, but also lays the groundwork for future research in RL methodologies. We believe \textit{HistoGym} will serve as a valuable tool for advancing the field of computational pathology.


\bibliographystyle{unsrt}  
\bibliography{ref.bib}

\section*{Appendix}

\subsection*{class HistoEnv(gym.Env)}

Custom environment for histology image analysis.

\subsubsection*{Parameters:}

\begin{itemize}
    \item \texttt{img\_path (str)}: Path to the histology image file.
    \item \texttt{xml\_path (str)}: Path to the XML file containing annotations.
    \item \texttt{tile\_size (int)}: Size of the tiles used for analysis.
    \item \texttt{result\_path (str)}: Path to save the resulting images.
\end{itemize}

\subsubsection*{Attributes:}

\begin{itemize}
    \item \texttt{UP (int)}: Action code for moving up.
    \item \texttt{DOWN (int)}: Action code for moving down.
    \item \texttt{LEFT (int)}: Action code for moving left.
    \item \texttt{RIGHT (int)}: Action code for moving right.
    \item \texttt{ZOOM\_IN (int)}: Action code for zooming in.
    \item \texttt{ZOOM\_OUT (int)}: Action code for zooming out.
    \item \texttt{STAY (int)}: Action code for staying in the same position.
    \item \texttt{img\_path (str)}: Path to the histology image file.
    \item \texttt{xml\_path (str)}: Path to the XML file containing annotations.
    \item \texttt{tile\_size (int)}: Size of the tiles used for analysis.
    \item \texttt{result\_path (str)}: Path to save the resulting images.
    \item \texttt{plt\_size (int)}: Size of the plot.
    \item \texttt{slide (openslide.OpenSlide)}: OpenSlide object for reading the histology image.
    \item \texttt{dz (DeepZoomGenerator)}: DeepZoomGenerator object for generating tiles.
    \item \texttt{dz\_level (int)}: Initial DeepZoom level.
    \item \texttt{OBS\_W (int)}: Observation width.
    \item \texttt{OBS\_H (int)}: Observation height.
    \item \texttt{STATE\_W (int)}: State width.
    \item \texttt{STATE\_H (int)}: State height.
    \item \texttt{coor\_xml (Coor)}: Coor object for parsing XML annotations.
    \item \texttt{coor\_dz\_all (Coor)}: Coor object for getting DeepZoom coordinates.
    \item \texttt{segment\_dz\_all (Coor)}: Coor object for getting segment coordinates.
    \item \texttt{if\_overlap (bool)}: Flag indicating if there is overlap.
    \item \texttt{overlap\_seg\_index (int)}: Index of the overlapping segment.
    \item \texttt{overlap\_ratio (float)}: Ratio of overlap.
    \item \texttt{n\_actions (int)}: Number of actions.
    \item \texttt{action\_space (gym.spaces.Discrete)}: Action space.
    \item \texttt{observation\_space (gym.spaces.Box)}: Observation space.
    \item \texttt{agent\_pos (list)}: Agent position in the form [z, x, y].
    \item \texttt{STATE\_D (int)}: Initial DeepZoom level for setting bounds.
    \item \texttt{state (numpy.ndarray)}: Current state.
    \item \texttt{count (int)}: Step count within the episode.
    \item \texttt{max\_step (int)}: Maximum number of steps per episode.
    \item \texttt{bound (list)}: List of bounds.
\end{itemize}

\subsubsection*{Methods:}

\begin{itemize}
    \item \texttt{\_\_init\_\_(self, img\_path, xml\_path, tile\_size, result\_path)}: Initializes the environment.
    \item \texttt{reset(self)}: Resets the environment and returns the initial state.
    \item \texttt{step(self, action)}: Takes a step in the environment based on the given action.
    \item \texttt{render(self, mode="save")}: Renders the current state of the environment.
\end{itemize}

\end{document}